\documentclass[a4paper,aps,preprintnumbers,twocolumn,amsmath,amssymb,prl]{revtex4}
\usepackage{bbm}
\usepackage{mathrsfs}
\usepackage{amsmath}
\usepackage[dvips]{graphicx}
\usepackage{epsfig}

\newcommand{\Ac}{{\cal A}}
\newcommand{\Bc}{{\cal B}}

\newcommand{\be}{\begin{equation}}
\newcommand{\ee}{\end{equation}}
\newcommand{\bea}{\begin{eqnarray}}
\newcommand{\eea}{\end{eqnarray}}
\newcommand {\ud}  {\mathrm{d}}
\renewcommand{\*}{\cdot}
\renewcommand{\vec}[1]{\mathbf{#1}}

\newcommand{\id}{\mathbbm{1}}
\newcommand{\tr}{\operatorname{Tr}}
\newcommand{\Vol}{\operatorname{Vol}}

\begin{document}
\title{Entanglement scaling in critical two-dimensional fermionic and bosonic systems}

\author{T. Barthel, M.-C. Chung, U. Schollw\"ock}
\affiliation{Institute for Theoretical Physics C, RWTH Aachen, D-52056 Aachen, Germany}

\begin{abstract}
We relate the reduced density matrices of quadratic bosonic and fermionic models to their Green's function matrices in a unified way and calculate the scaling of bipartite entanglement of finite systems in an infinite universe exactly. For critical fermionic 2D systems at $T=0$, two regimes of scaling are identified: generically, we find a logarithmic correction to the area law with a prefactor dependence on the chemical potential
that confirms earlier predictions based on the Widom conjecture. If, however, the Fermi surface of the critical system is zero-dimensional, we find an area law with a sublogarithmic correction. For a critical bosonic 2D array of coupled oscillators at $T=0$, our results show that entanglement follows the area law without corrections.
\end{abstract}

\date{\today}

\maketitle

Entanglement is a key feature of the non-classical nature of quantum mechanics. It is a necessary resource for quantum computation and is at the heart of interesting connections between quantum information theory and traditional quantum many-body theory, such as in quantum critical phenomena \cite{VidalJ2004,VidalG2003,Osborne2002}
or the quantum Hall effect \cite{Samuelsson2004,Kim2004}.

One of the most widely used entanglement measures is the entropy of bipartite entanglement, which is nothing but the von Neumann entropy of quantum statistics: For a pure state $|\Psi_\mathcal{AB}\rangle$ of a bipartite "universe" $\Ac\Bc$ consisting of system ${\Ac}$ and environment ${\Bc}$ it is given by 
    $S_{\Ac} =  - \tr{\rho_{\Ac} \log_2 \rho_{\Ac}}$, 
where $\rho_{\Ac} = \tr_{\Bc}{|\Psi_{\Ac\Bc}\rangle \langle \Psi_{\Ac\Bc}|}$ is the reduced density matrix of system ${\Ac}$.
 
An important question to ask is how entanglement entropy scales with the size 
of the system, assuming the universe to be in the thermodynamic limit. This was first studied by Beckenstein in the context of black hole entropy \cite{Beckenstein}. As opposed to thermodynamic entropy, which is extensive, entanglement entropy was found to be proportional to the area of the black hole's event horizon, its physical locus being essentially the hypersurface separating system and environment. Entanglement entropy scaling hence depends decisively on the dimension $d$ of the universe.

This observation has given rise to a long string of studies of this so-called area law.  
In one dimension $d=1$, scaling is well understood both for fermions [2,\,\,7--11] and bosons \cite{Werner,Skrovesth}. For one-dimensional spin chains, one finds that the entanglement entropy $S_{\Ac}(L)$ of a system $\Ac$ of linear size $L$ saturates away from criticality, but scales  as $\log_2 L$ at criticality \cite{VidalG2003}. In the latter case, conformal field theory (CFT) yields \cite{Korepin2,Cardy} $S_{\Ac}(L) = \frac{c+\bar{c}}{6} \log_2{L} + k$,    
where $c$ and $\bar{c}$ are the holomorphic and the anti-holo\-mor\-phic central charges of the field
theory. Essentially, there is no physical limit to the boundary region between system and environment. 
      
The situation is far less clear in higher dimensions $d>1$. The area law implies that the entanglement away from criticality is essentially proportional to the surface area of system $\Ac$
\be \label{eqn:arealaw}
      S_{\Ac} \sim L^{d-1}\,, 
\ee
as confirmed in analytical calculations for non-critical bosonic coupled oscillators \cite{Cramer}.
  
At criticality, the correlation length diverges and one may expect corrections to the area law, as for $d=1$. For critical ground states of fermionic tight-binding Hamiltonians entanglement was indeed found to scale as
\be \label{eqn:surfacearea}
      S_{\Ac} \sim L^{d-1} \log_2 L\,, 
\ee
for both lattice models \cite{Wolf} and continuous fields \cite{Klich}. The prefactor could only be derived \cite{Klich} assuming (i) the validity of the Widom conjecture \cite{Widom1981} and (ii) its applicability to the functional form of binary entropy. For bosons at criticality, numerical evidence for the area law \eqref{eqn:arealaw} was found for a three-dimensional array of coupled oscillators \cite{Srednicki}. Callan and Wilczek derived the area law in approximative field theoretical calculations \cite{Callan}.

Beyond the fundamental physical interest, entanglement scaling sets the scope of entanglement-based numerical methods such as the density-matrix renormalization-group (DMRG) \cite{White1992-11, Schollwoeck2005}, as the computation time required to simulate a quantum state using these methods on classical computers increases exponentially with its entanglement entropy.

In this Letter, we study the bipartite entanglement in a {\em unified} treatment of a class of {\em exactly solvable} two-dimensional fermionic and bosonic models at $T=0$. To this purpose, we relate the reduced density matrix of a quadratic model to its Green's function matrices, generalizing work by Cheong and Henley \cite{CheongHenley} based on a coherent-state method developed by Chung and Peschel \cite{Chung}. For the critical fermionic two-dimensional tight-binding model we find as expected \eqref{eqn:surfacearea}, but our exact calculation allows to identify the dependence of the scaling law prefactor on the chemical potential $\mu$. We exactly verify the behavior predicted in \cite{Klich}, where the validity of the Widom conjecture and its applicability to the binary entropy were assumed. Interestingly, we observe a {\em sublogarithmic} correction to the area law if the gap of the model closes in a zero-dimensional region of momentum space (i.e.\ one or more points). For a critical bosonic two-dimensional model of coupled harmonic oscillators we find the entanglement to saturate to the area law \eqref{eqn:surfacearea}, which confirms \cite{Srednicki,Callan}.
      
The generic quadratic Hamiltonians studied here are
       \be \label{eqn:Hfermion}
        H_{F,B} = \sum_{ij} \left[ a_i^{\dagger} A_{ij}a_j + \frac{1}{2} 
         (a_i^{\dagger} B_{ij} a_j^{\dagger} + h.c. )\right]\,,
      \ee  
where $a_i \equiv c_i$ and $a_i \equiv b_i$ are fermionic and bosonic operators for $H_F$ and $H_B$ respectively.

\emph{Calculating entanglement from Green's matrices.}--- We consider a bipartite universe $\Ac\Bc$ of $N$ levels (or sites). System $\Ac$ consists of $n$ sites;
in our calculations we will eventually take the thermodynamic limit $N\rightarrow\infty$. 
The relation between the Green's function matrices of system $\Ac$ and its reduced density matrix $\rho_{\Ac} = \tr_{\Bc}{\rho}$ can be derived by determining the matrix elements of the full density matrix $\rho$ with respect to coherent states and integrating out the variables of the environment ${\Bc}$. 
Here, we focus on the key steps and results of this method. A detailed derivation will be given elsewhere.

{\emph{(a) Fermionic systems.}}
The block Green's function matrix with respect to the operators $A_i = c^{\dagger}_i + c_i$ and $B_i =  c^{\dagger}_i-c_i$, as defined by
\be \label{eqn:GreenMF}
   [{G}_{BA}]_{ij} = \tr{ \rho B_i \,A_j}\quad \text{with}\quad i,j\in \Ac,
\ee
can be obtained \emph{exactly} for the solvable Hamiltonian $H_F$, Eq.~\eqref{eqn:Hfermion}, following \cite{LSM}. It can then be shown that $\rho_{\Ac}$ is given by 
\be 
\label{eqn:MEforGM}
  \begin{split}
  \langle \boldsymbol{\xi} | \rho_{\Ac} |\boldsymbol{\xi}'\rangle & = 
   \det{\frac{1}{2}(\id  - {G}_{BA})} \\
   &\times e^{ -\frac{1}{2} (\boldsymbol{\xi}^{\star}-\boldsymbol{\xi}')^T 
      \cdot({G}_{BA}+ \id )( {G}_{BA}- \id )^{-1}\cdot
    (\boldsymbol{\xi}^{\star}+{\boldsymbol{\xi}}')}\,,
  \end{split}
\ee
where $\boldsymbol{\xi} =  \{\xi_1\,\cdots\xi_n\}$ are the Grassmann
variables associated with system ${\Ac}$, and $|\xi\rangle$ are the corresponding coherent states with  
$c_i|\boldsymbol{\xi}\rangle = \xi_i |\boldsymbol{\xi}\rangle$.

To calculate the (entanglement) entropy of system $\Ac$, we diagonalize $\rho_{\Ac}$  by the Bogoliubov transformation 
\be \label{eqn:fq}\textstyle
 f_q  = \sum_{i}\left[ \frac{P_{qi}+Q_{qi}}{2} \;c_i +
 \frac{P_{qi}-Q_{qi}}{2}
 \;c_i^{\dagger} \right]\,,
\ee 
where ${P} {P}^T = {Q} {Q}^{T} = \id$ (due to the anti-commutation rules), 
 $P_q \;G_{BA}^T =  \nu_q Q_q$ and  $Q_q \;G_{BA} =  \nu_q P_q$\,.
The diagonalized reduced density matrix reads
\be \label{eqn:DiagRDM}\textstyle
 \rho_{\Ac} = \left(\prod_{q} \frac{1-\nu_q}{2}\right)\cdot
 e^{-\sum_{q} \varepsilon_q 
f_q^{\dagger} f_q} , 
\ee 
with pseudo-energies $\varepsilon_q = \ln \frac{1-\nu_q}{1+\nu_q}\,$, such that the entanglement entropy reads  
\be\label{eqn:FEntropy}\textstyle
  S_{\Ac} = \sum_{q=1}^{n} h(\frac{1+\nu_q}{2}),
\ee
\be\label{eqn:binaryEntropy} 
\quad h(x) = -x \log_2 x - (1-x) \log_2 (1-x)\,\ee
being the binary entropy.

\emph{(b) Bosonic systems.}
For the quadratic Hamiltonian $H_B$ the block Green's function matrices $G_{AA}$ and $G_{BB}$  with respect to the operators $A_i = b^{\dagger}_i + b_i$ and $B_i =  b^{\dagger}_i-b_i$ can be obtained as  in \cite{Colpa1978-93}. With respect to the bosonic coherent states 
$b_i |{\boldsymbol \phi} \rangle = \phi_i |{\boldsymbol \phi} 
\rangle$, the reduced density matrix then reads 
\be
\begin{split}
  \langle \boldsymbol{\phi}| \rho_{\Ac}| \boldsymbol{\phi}' \rangle = &
K' e^{\frac{1}{4} (\boldsymbol{\phi}^{\star} + \boldsymbol{\phi}')^T\cdot 
(G_{AA} -\id )(G_{AA} + \id )^{-1}\* 
 (\boldsymbol{\phi}^{\star} +  \boldsymbol{\phi}')} \\ \times
 & e^{ -\frac{1}{4} (\boldsymbol{\phi}^{\star} -  \boldsymbol{\phi}')^T\* 
  (G_{BB}-\id )^{-1}(G_{BB}+\id )\*
 (\boldsymbol{\phi}^{\star} -  \boldsymbol{\phi}')}\,, 
  \end{split}
\ee
where $K' = \sqrt{\det{(1+G_{AA})(1-G_{BB})}} $ is determined by the normalization  of $\rho_{\Ac}$.

The Bogoliubov transformation
\be \label{eqn:gq}\textstyle
 g_q  = \sum_{i}\left[ \frac{P_{qi}+Q_{qi}}{2} \;b_i +
 \frac{P_{qi}-Q_{qi}}{2}
 \;b_i^{\dagger} \right], 
\ee
with $P^T Q = Q^T P = \id $, $P_q G_{AA}  = \mu_q Q_q$ and $Q_q G_{BB}  = - \mu_q P_q$
diagonalizes $\rho_{\Ac}$, giving
\be\label{eqn:BBDMEV}\textstyle
   \rho_{\Ac} = \left(\prod_q \frac{2}{\mu_q+1}\right)
   e^{-\sum_q \epsilon_q g_q^{\dagger} g_q }\,,
\ee 
where  $\epsilon_q = \ln\left(\frac{\mu_q+1}{\mu_q-1}\right)$ are pseudo-energies. The entanglement entropy $S_{\Ac}$ is the sum of the quasi-particle entropies, 
\be \label{eqn:BEntropy}
  \textstyle
    S_{\Ac}  = \sum_{q=1}^n \left(  \frac{\mu_q+1}{2} \log_2 \frac{\mu_q+1}{2}\, -\,  
    \frac{\mu_q-1}{2}
              \log_2 \frac{\mu_q-1}{2}\right)\,.
\ee

\emph{Critical fermionic entanglement and the Widom conjecture.}---
The form of the logarithmic correction to the entanglement in $d>1$ dimensional critical fermion models and bounds on it have been derived by Wolf \cite{Wolf}, Gioev and Klich \cite{Klich}. Assuming that the Widom conjecture \cite{Widom1981} holds also for $d>1$ and that the non-analyticity of the binary entropy $h$ can be ignored at one point in the calculation, Gioev and Klich \cite{Klich} arrive at
\be\label{eqn:S2dfermion}
    S_{\Ac} \equiv S_{\Omega}(L) = c(\mu) L \log_2 L+o(L \log_2L),
\ee 
\be\label{eqn:c-mu-2dfermion}
c(\mu)=\frac{1}{2\pi}\frac{1}{12}\int_{\partial\Omega}\ud S_x\int_{\partial \Gamma(\mu)}\ud S_k\,|\vec{n}_\vec{ x}\cdot\vec{n}_\vec{k}|\,,
\ee
where $\Omega$ is the real-space region of $\Ac$, rescaled by $L$ such that $\Vol(\Omega)=1$. Vectors
$\vec{n}_\vec{x}$ and $\vec{n}_\vec{k}$ denote the normal vectors on the surface $\partial\Omega$ and the Fermi surface $\partial\Gamma(\mu)$.
With the method introduced above, one can calculate the entanglement for finite $L$ \emph{exactly} and thus check \eqref{eqn:S2dfermion}, also shedding some light on the validity of the assumptions leading to \eqref{eqn:c-mu-2dfermion}.

\begin{figure}[t]
\begin{center}
\epsfig{file=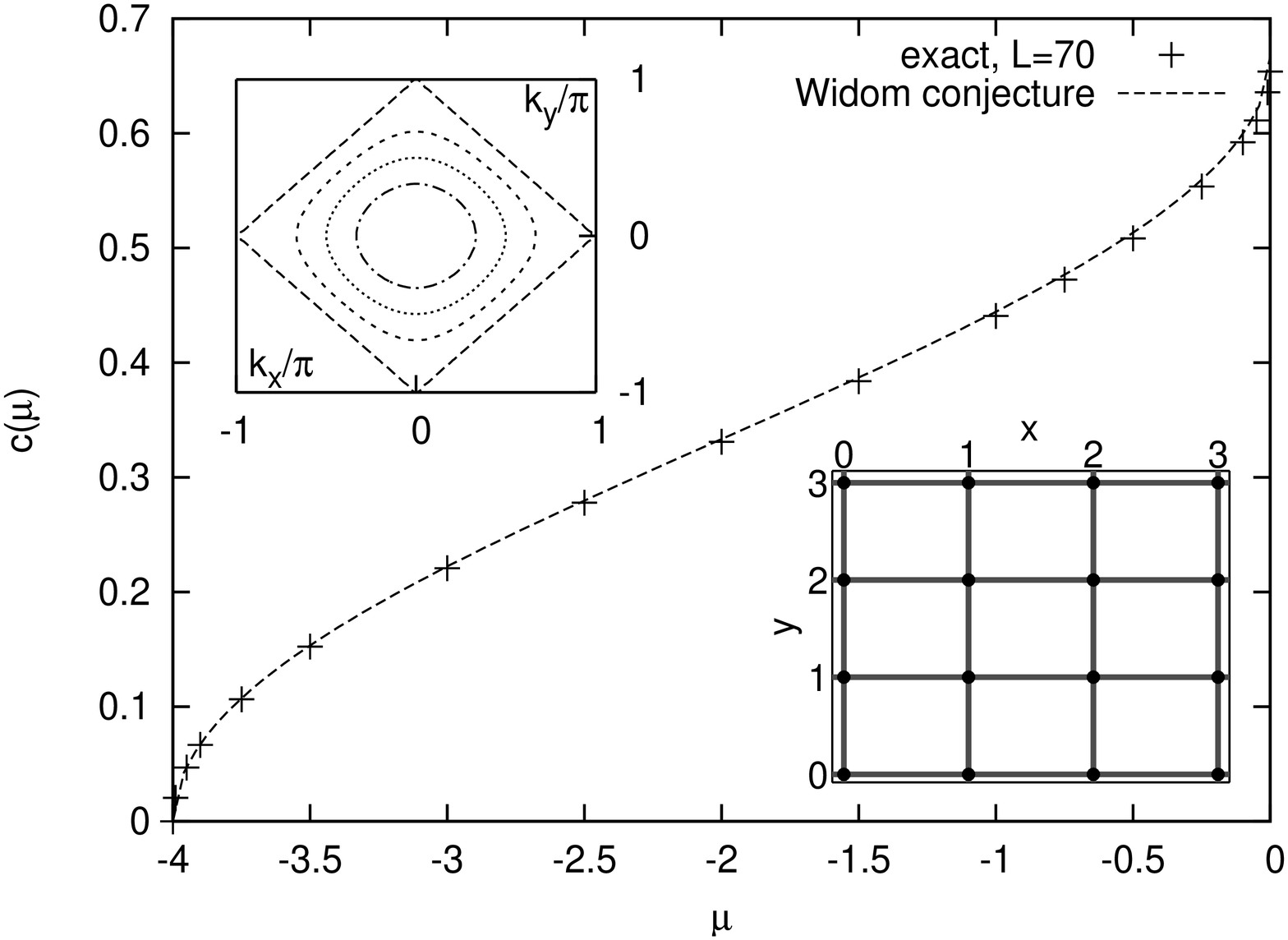,angle=0,width=1\linewidth}
\end{center}\vspace{-1em}
\caption {
The prefactor $c(\mu)$ in the entanglement scaling law as a function of the chemical 
potential $\mu$ for the  ground state of the two-dimensional fermionic tight-binding model in comparison to the 
result of Gioev and Klich \cite{Klich}. Insets show the hopping parameters and the Fermi surfaces for $\mu= -3,-2,-1,0$.} \label{fig:cmu}
\end{figure}

\begin{figure}[t]
\begin{center}
\epsfig{file=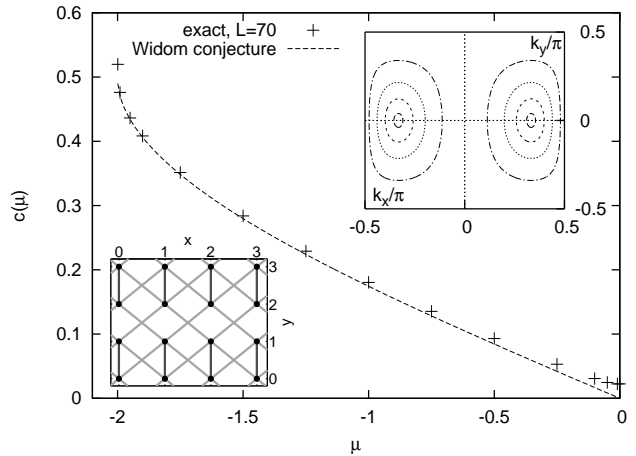,angle=0,width=1\linewidth}
\end{center}\vspace{-1em}
\caption {The scaling prefactor $c(\mu)$ for the  ground state of a two-dimensional fermionic tight-binding model with next-nearest neighbor hoppings in comparison to the result of \cite{Klich}.  Insets show the hopping parameters and the Fermi surfaces for $\mu\in [-0.25,-1.75]$ in the quartered Brillouin zone.} \label{fig:cmu2}
\end{figure}

The dispersion relation of the two-dimensional tight-binding model with periodic boundary conditions
\be\label{eqn:H2DF1}
\textstyle
H=-\sum_{x,y}\left(c^\dagger_{x,y}c_{x+1,y}+c^\dagger_{x,y}c_{x,y+1}+h.c.\right)
\ee
is $E(\vec{k})=-2\cdot\left(\cos k_x+\cos k_y\right)$. The ground state Green's function matrix, from which we calculate the entanglement, reads in the thermodynamic limit
\be\textstyle
 G_{\vec{r},\vec{r}'} =  \int_{\Gamma(\mu)} \frac{\ud^2 k}{(2\pi)^2}\, 
 e^{i \vec{k} \cdot (\vec{r} - \vec{r}')}\,,
 \ee
with $\vec{r}=(x,y)$. 
Fig.~\ref{fig:cmu} shows the scaling prefactor $c(\mu)$ as fitted from the exact entanglement of an $L\times L$ square with the rest of the universe, which was obtained from \eqref{eqn:FEntropy}. It is in excellent agreement with \eqref{eqn:c-mu-2dfermion} and supports thus the Widom conjecture for $d=2$. The same agreement was found in the model
\be\label{eqn:H2DF2}
\begin{split}\textstyle
H=-\sum_{x,y}&\big((1+(-1)^y)c^\dagger_{x,y}c_{x,y+1}\\\textstyle
&+c^\dagger_{x,y}c_{x+1,y+1}+c^\dagger_{x,y}c_{x-1,y+1}+h.c.\big)
\end{split}
\ee
which has a two-banded dispersion relation 
$E(\vec{k})=\pm2\sqrt{1+4 \cos k_x\cos^2 k_y+4\cos^2 k_x\cos^2 k_y}$
and a disconnected Fermi surface for $\mu\in[-2,2]$, Fig.~\ref{fig:cmu2}.
\begin{figure}[t]
\begin{center}
\epsfig{file=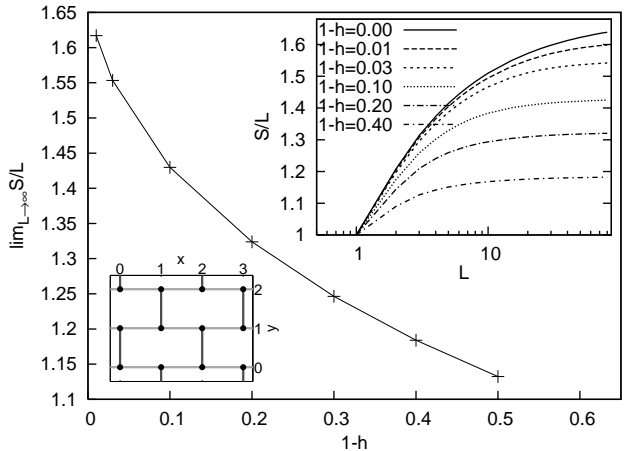,angle=0,width=1\linewidth}
\end{center}\vspace{-1em}
\caption {The upper right inset shows the entanglement entropy per surface unit $S_\Omega(L)/L$ (block of $n=L^2$ sites) for a two-dimensional fermionic tight-binding model with modulated vertical hopping (see the second inset) at $T=0$. The energy gap $4(1-h)$ closes in the point $\vec{k}=(\pi,0)$. The extrapolation $\lim _{L\to\infty} S_\Omega(L)/L$ suggests a divergence for $h\to 1$.} \label{fig:cmuPointGap} 
\end{figure}

Especially for a comparison with bosonic systems, it is interesting to investigate models with a zero-dimensional Fermi surface. In particular we choose the two-dimensional model 
\be\label{eqn:H2DFpointGap}
\begin{split}\textstyle
H=-\sum_{x,y}\big(&h\*c^\dagger_{x,y}c_{x+1,y}\\
&+(1+(-1)^{x+y})c^\dagger_{x,y}c_{x,y+1}+h.c.\big)\,,
\end{split}
\ee
which has for $0\leq h\leq 1$ the two-band dispersion relation $E(\vec{k})=\pm 2\sqrt{1+h^2\cos^2k_x+2h\cos k_x\cos k_y}$, i.e.\ a gap of size $4(1-h)$ at $\vec{k}=(\pi,0)$. 
Fig.~\ref{fig:cmuPointGap} shows for $\mu=0$ and $h\to 1$ how the entanglement converges to the area law with a \emph{sublogarithmic} correction, $S_\Omega(L)=L\*o(\log_2L)$, meaning $\lim_{L\to\infty}S_\Omega(L)/(L\log_2L)=0$.
The curves $S_\Omega(L)/L$ for finite gaps were extrapolated to obtain $\lim_{L\to\infty} S_\Omega(L)/L$. Those values indicate indeed a divergence for $h\to 1$.
This result is consistent with Eq.~\eqref{eqn:S2dfermion}, as the scaling coefficient $c(\mu)$, Eq.~\eqref{eqn:c-mu-2dfermion}, vanishes for systems in $d>1$ dimensions with a zero-dimensional Fermi surface.
 Further investigations have to determine the analytical form of the sublogarithmic correction and its universality.

\emph{Critical bosonic entanglement.}--- 
An important question is whether the logarithmic correction observed in the entanglement scaling law for critical one-dimensional bosonic systems is also present in higher dimensional systems.
 To investigate this, we examine a two-dimensional system of coupled 
oscillators
\be\label{eqn:HharmLattice}
\begin{split}\textstyle
     H = \frac{1}{2}&\textstyle\sum_{x,y}\,\big(\Pi_{x,y}^2 
    + \omega_0^2 \,\Phi_{x,y}^2\\\textstyle
    &+  (\Phi_{x,y}-\Phi_{x+1,y})^2 + (\Phi_{x,y}-\Phi_{x,y+1})^2\big),   
\end{split}
\ee 
where $\Phi_{x,y}$, $\Pi_{x,y}$ and $\omega_0$ are coordinate, momentum 
and self-frequency of the oscillator at site $\vec{r}=(x,y)$. The masses and coupling strengths are set to unity. The system has the dispersion relation $E(\vec{k}) = \sqrt{\omega_0^2 + 
4\sin^2{k_x/2}+4\sin^2{k_y/2}}$, i.e.\ a gap $\omega_0$ at $\vec{k}=\vec{0}$. 

In the low-energy limit, the harmonic oscillators can be reduced to a field theory only containing $(\nabla\phi)^2$, which describes a massless free bosonic model. The scaling of entanglement in this model has been studied by Srednicki \cite{Srednicki} numerically in $d=3$ dimensions and by Callan and Wilczek \cite{Callan} with approximate field theoretical methods for all $d>1$. Both provide evidence for the area law \eqref{eqn:arealaw}.
       
\begin{figure}[h]
\begin{center}
\epsfig{file=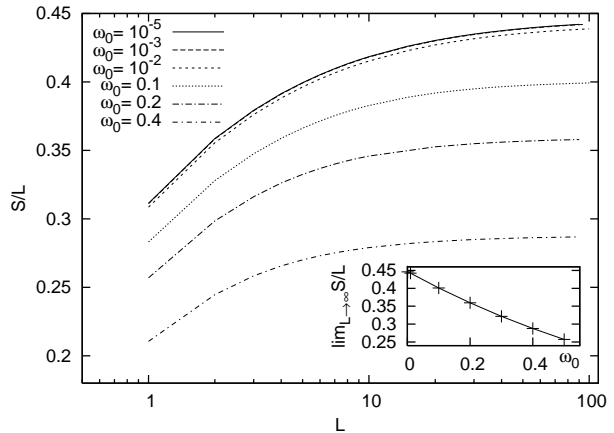,angle=0,width=1\linewidth}
\end{center}\vspace{-1.5em}
\caption{
The entanglement entropy per surface unit $S_\Omega(L)/L$ (block with $n=L^2$ sites) for the bosonic two-dimensional system of coupled harmonic oscillators at $T=0$. The curves converge for small energy gaps $\omega_0$. The extrapolation $\lim _{L\to\infty} S_\Omega(L)/L$ yields, for the critical limit, the area law $S_\Omega(L)\approx0.45\* L$.}  \label{fig:2DHO} 
\end{figure}

Applying the transformation $b_{\boldsymbol i} = \sqrt{\frac{\omega}{2}}\* (\Phi_{\boldsymbol i} 
   + \frac{i}{\omega} \Pi_{\boldsymbol i})$ with $\omega  = \sqrt{\omega_0^2 + 4}$,
the Hamiltonian \eqref{eqn:HharmLattice} is mapped to the canonical form \eqref{eqn:Hfermion} and is thus amenable to the method introduced above. The translationally invariant block Green's function matrices $G_{AA}$ and $G_{BB}$ are for $T=0$
\be \label{eqn:GAAGBBHO}\nonumber
      \begin{split}
      G_{AA}(\vec{r},\vec{0})  &\textstyle =  \frac{1}{\pi^2} 
        \int_{0}^{\pi} \int_{0}^{\pi} \ud^2 k  \,\frac{\omega}{E(\vec{k})}
      \cos{k_x x} \cos{k_y y} \\
      G_{BB}(\vec{r},\vec{0}) &\textstyle=   -\frac{1}{\pi^2}
         \int_{0}^{\pi} \int_{0}^{\pi} \ud^2 k \,\frac{E(\vec{k})}{\omega}
      \cos{k_x x \cos{k_y y}}\, 
      \end{split}
\ee
and the entanglement is obtained from Eq.~\eqref{eqn:BEntropy}.
Special care has to be taken for the limit $\omega_0\to 0$, as this results in a singularity of the integrand for $G_{AA}$.
Fig.~\ref{fig:2DHO} displays the entanglement entropy as a function of the linear block size $L$ for several $\omega_0$. The curves converge for $\omega_0\to 0$ and a finite-size scaling analysis yields $\lim_{\omega_0\to 0}\lim_{L\to\infty}S_\Omega(L)/L \approx
0.45$, i.e.\ the critical model obeys for $d=2$ the area law $S_\Omega(L)\approx 0.45\*L$.

\emph{Conclusions.}---
A relation between Green's function matrices of quadratic fermionic and bosonic Hamiltonians to reduced density matrices was used to study bipartite entanglement in critical 2D systems. We identified and presented exact quantitative results for three different regimes of entanglement scaling.
Those findings demonstrate the subtle nature of entanglement at criticality, the physical explanation of which remains a challenging topic for future research.  

This work was supported by the DFG. A critical reading of an early stage of this manuscript by H.-J. Briegel and E. Rico is gratefully acknowledged.

\end{document}